\documentclass{ws-procs9x6}
\newcommand{\bqa}{\begin{eqnarray}}
\newcommand{\eqa}{\end{eqnarray}}
\newcommand{\bc}{\begin{center}}
\newcommand{\ec}{\end{center}}

\newcommand{\ga}{\gamma}
\newcommand{\si}{\sigma}
\newcommand{\ta}{\tau}

\newcommand{\rh}{\rho}
\newcommand{\om}{\omega}
\newcommand{\unit}{1\hspace*{-1.5mm}\mbox{l}}
\begin{document}

\title{Thermal dilepton rates and in-medium\\
hadron properties from lattice QCD}

\author{I. Wetzorke}

\address{NIC/DESY Zeuthen, \\ 
Platanenallee 6, \\
D-15738 Zeuthen, Germany\\
E-mail: Ines.Wetzorke@desy.de}  

\maketitle

\abstracts{
Recent progress of lattice investigations in thermal physics is summarized in
this contribution. Hadronic spectral functions can be reconstructed from
correlation functions in Euclidean time based on the Maximum Entropy Method
without a priori assumptions on the spectral shape. The thermal modifications
of hadron properties are investigated in the scalar, pseudo-scalar, vector and
axial-vector channels near and above the deconfinement transition temperature
for charmonium systems as well as for light quarks. Moreover, the reconstructed
vector meson spectral function allows to extract the thermal cross section for
the production of dilepton pairs at vanishing momentum.}

\section{Introduction}
Precise lattice QCD calculations of hadron masses and decay constants
at zero temperature were performed in the recent years. A challenge is
the investigation of modifications of hadronic properties at finite
temperature, in particular the changes of the spectral shape.
The spectral functions are directly relevant for experimental annihilation
cross sections in relativistic heavy-ion collisions, e.g. the temperature
dependence of the vector meson mass and width is related to changes of the
thermal dilepton spectrum.

On the lattice the particle masses are usually obtained from the exponential
decrease of the correlation function at zero momentum at large Euclidean times
$\ta$:
\bqa
G(\ta) = \int d^3 x \;
\langle\; J_H(\vec{x},\ta) J_H^\dag(\vec{0},0)\;\rangle
\sim e^{-m \; \ta},
\eqa
where the hadronic current
$J_H (\tau, \vec{x}) = \overline{\psi} (\tau, \vec{x}) \Gamma
\psi (\tau, \vec{x})$ is given in the scalar, pseudo-scalar, vector
or axial-vector channel with $\Gamma \in [\unit,\ga_5,\ga_\mu,\ga_\mu \ga_5]$.
At finite temperature $T$ the Euclidean time extent is restricted by $1/T$
rendering a reliable determination of the particle masses increasingly
difficult with rising temperature.

Another possibility to extract the masses and decay constants is provided
by the hadronic spectral spectral functions $\si_H$, which are related to the
correlation functions in Euclidean time by
\bqa
G(\tau,\vec{p}) = \int {\rm d^3 x} \; e^{i \vec{p} \vec{x}} \;
\langle J_H (\tau, \vec{x}) J_H^\dag (0, \vec{0}) \rangle
= \int\limits_0^\infty \; {\rm d}\om \; K(\ta,\om) \; \si_H(\om,\vec{p},T),
\eqa
where $K$ is the integration kernel in the continuum
\bqa
K^{cont}(\ta, \om)=\frac{\cosh(\om(\ta - 1/2T))}{\sinh(\om/2T)}.
\label{cont}
\eqa
The reconstruction of a smooth spectral function given only a limited
number of points of the correlator in Euclidean time can be classified as
typical ill-posed problem. Previous analyses have been done under restricting
assumptions on the spectral shape. The modern solution is provided by the 
Maximum Entropy Method (MEM), which has been applied successfully to many
similar problems in physics \cite{Jarrel}. This approach has been first applied
in lattice QCD by \cite{Hatsuda}. It allows to obtain the most probable
spectral function without a priori assumptions on the spectral shape by
maximizing the prior probability in addition to usual least-square
minimization.

\section{Meson spectral functions at $T=0$ and $T=\infty$}
\begin{figure}[ht]~\\
\epsfxsize=5.4cm\epsfysize=3.6cm\epsfbox{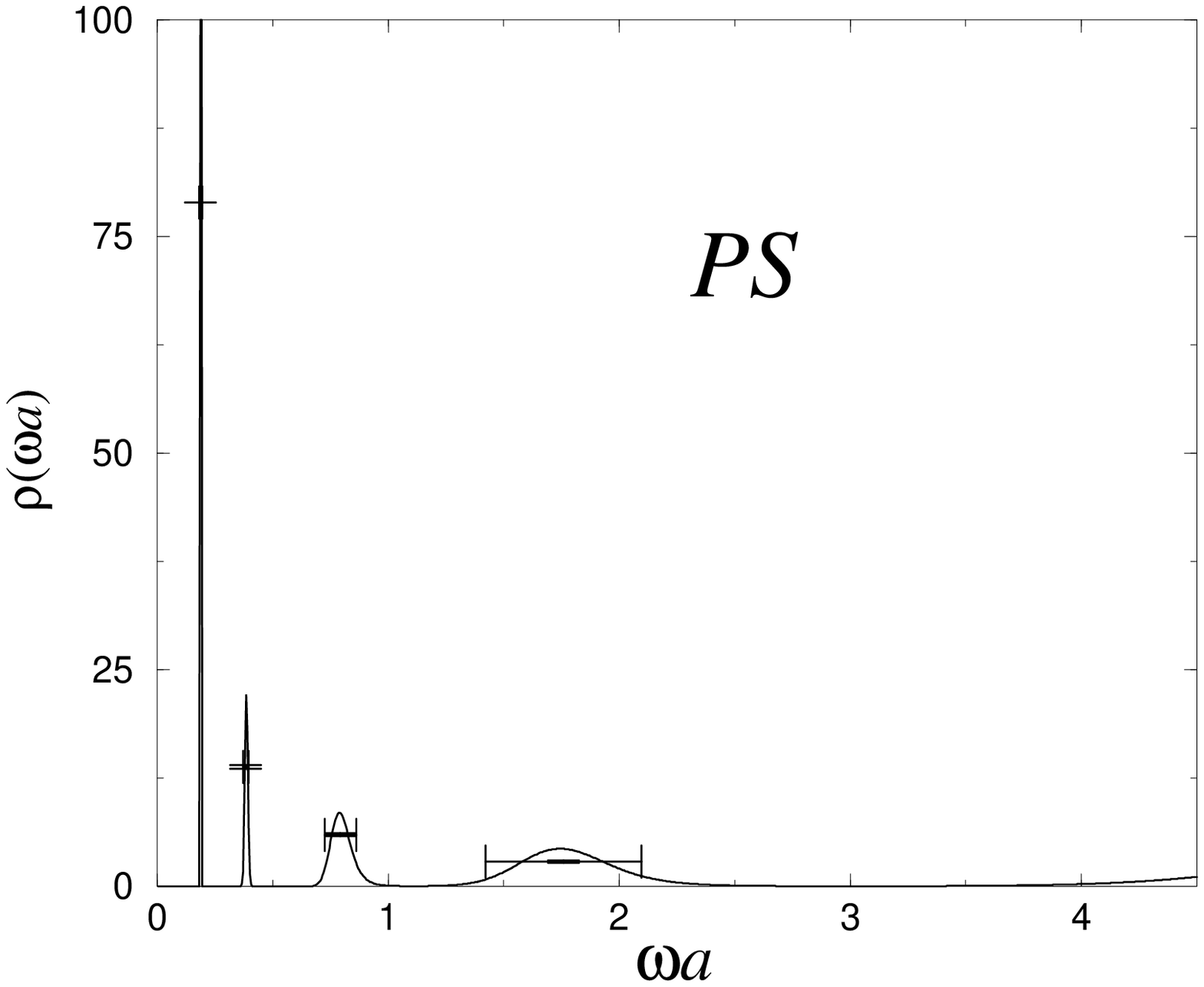}
\hspace*{3mm}                      
\epsfxsize=5.4cm\epsfysize=3.6cm\epsfbox{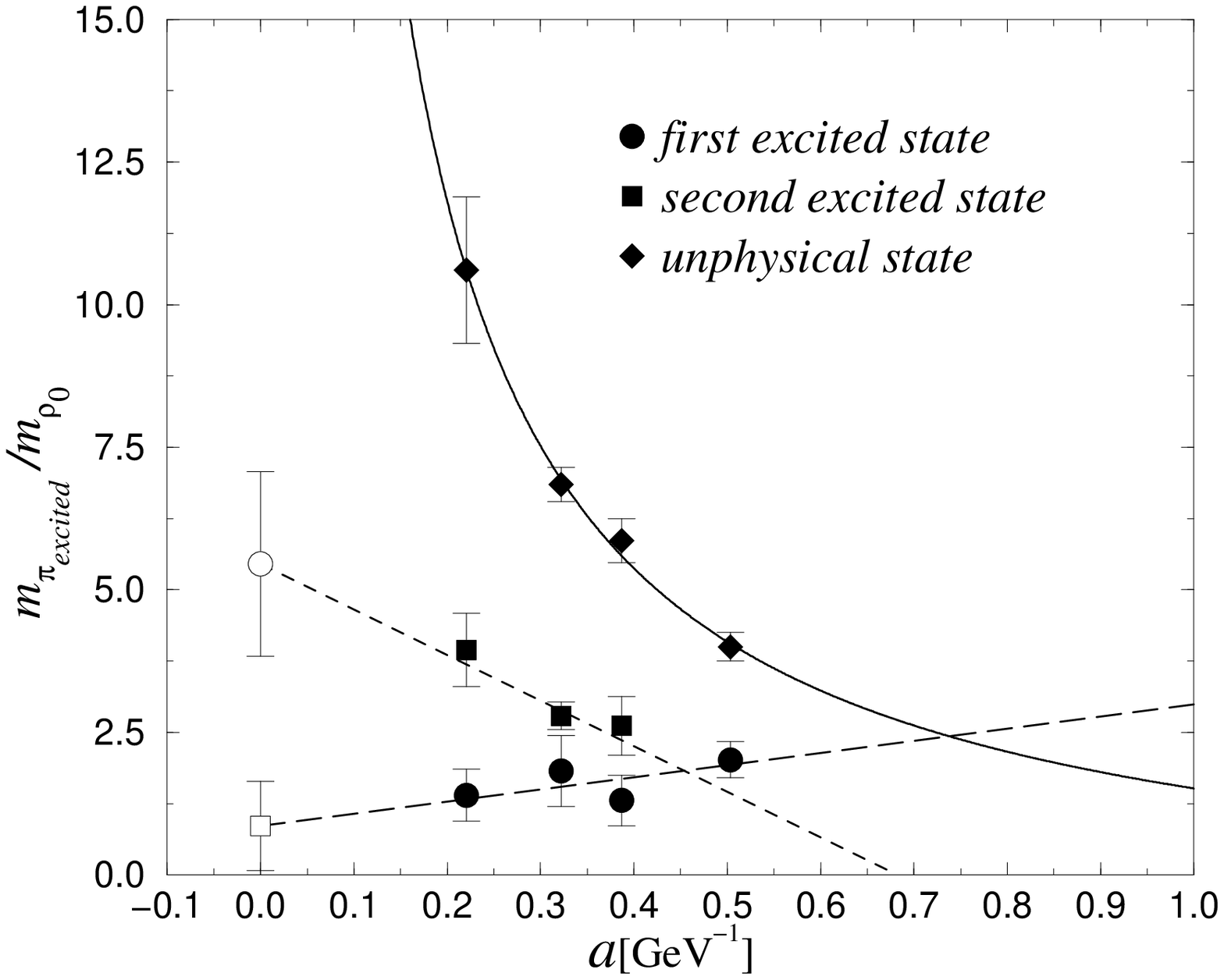}   
\caption{Pseudo-scalar spectral function at $T=0$ (left) and continuum
  extrapolation (right) taken from \protect\cite{Yamazaki}, results for
  the vector meson are similar.}
\label{fig_T=0}
\end{figure}

Before we proceed to hadronic spectral functions at finite temperature,
where very little is known about the spectral shape so far,
the behavior in the limiting cases $T=0$ and $T=\infty$ should be
investigated. At zero temperature $\delta-$function like peaks
are expected for the ground and excited states at low energies,
while the meson spectral functions should be proportional to $\om^2$ at high
energies. The precise quenched QCD data of the CP-PACS Collaboration
has been used for a detailed MEM analysis of ground and excited state meson masses
as well as decay constants \cite{Yamazaki}. They have investigated pseudo-scalar
and vector meson correlators on very large lattices ($32^3 \times 56$ to $64^3
\times 112$) to at four different values of the lattice spacing $a$.
The pseudo-scalar spectral function $\rh(\om)=\si_{PS}/\om^2$ in figure
\ref{fig_T=0} shows several peaks for the ground and excited states.
On smaller lattices only the ground state \cite{Wetzorke} and one excited state
\cite{Hatsuda} can be resolved by the reconstructed spectral function.
In general the peaks are broader and reduced in height for smaller lattice
sizes and lower statistics.
 
The ground state masses extracted from the spectral functions are the same as
the ones obtained from exponential fits of the correlator, while the mass of
the first excited state could be determined much more precisely than before.
The broad state slightly below 
$\om a = 2$, similarly observed in \cite{Wetzorke,Hatsuda}, diverges in the
continuum limit (fig. \ref{fig_T=0} right) and thus has been identified as
unphysical lattice artifact (cut-off effect) of the heavy fermion doublers
appearing in the Wilson type fermion formulations.

At finite temperature excited states and continuum-like contributions
are expected to gain in influence on the spectral shape. The spectral functions
at $T=\infty$ for free massless quarks are known from leading order
perturbation theory, e.g. $\si_{PS}^{free} = 3/8 \pi^{-2} \om^2 \tanh(\om/4T)$.
It is mandatory to verify that MEM allows to reconstruct such a continuum-like
shape correctly. It turned out \cite{lat01} that this shape cannot be obtained
using the continuum integration kernel of eq.(\ref{cont}) due to large
contributions of heavy Wilson fermion doublers at high energies. The
application of the finite lattice approximation of the continuum kernel
\bqa
K^{lat}(\ta,\om)= 2\om / N_\ta \; \sum_{n=0}^{N_\ta-1}
\frac{\exp(-i \; \om_n \ta)}{4\; \sin^2(\om_n/2) + \om^2}
\eqa
absorbs the cut-off effects efficiently and
allows an almost perfect reconstruction already for 8-16 points in Euclidean
time. The MEM analysis reported in the following sections is therefore
performed using this lattice adapted integration kernel.

In a very recent paper \cite{infinite} the authors studied
the cut-off effects in detail in the infinite temperature limit, where the
spectral functions can be calculated analytically \cite{free}. They found
considerable cut-off effects for the usual Wilson fermion formulation, which
are strongly suppressed and shifted to the very high energy regime using a
truncated perfect fermion action. 

\section{Medium modifications for light mesons}
The thermal changes of meson spectral functions in the light quark sector
have been investigated in the range 0.4-3.0 times the chiral phase transition
temperature. The simulations were performed in quenched QCD with a
non-perturbatively improved fermion action on large isotropic lattices
up to $64^3 \times 24$.
Calculations below $T_c$ were carried out at non-vanishing quark mass
corresponding to a pion mass of about 500 MeV. Above the critical temperature
the simulations could be performed directly at approximately zero quark mass,
since massless Goldstone modes are no longer present.
\begin{figure}[ht]~\\
\epsfxsize=5.6cm\epsfysize=3.6cm\epsfbox{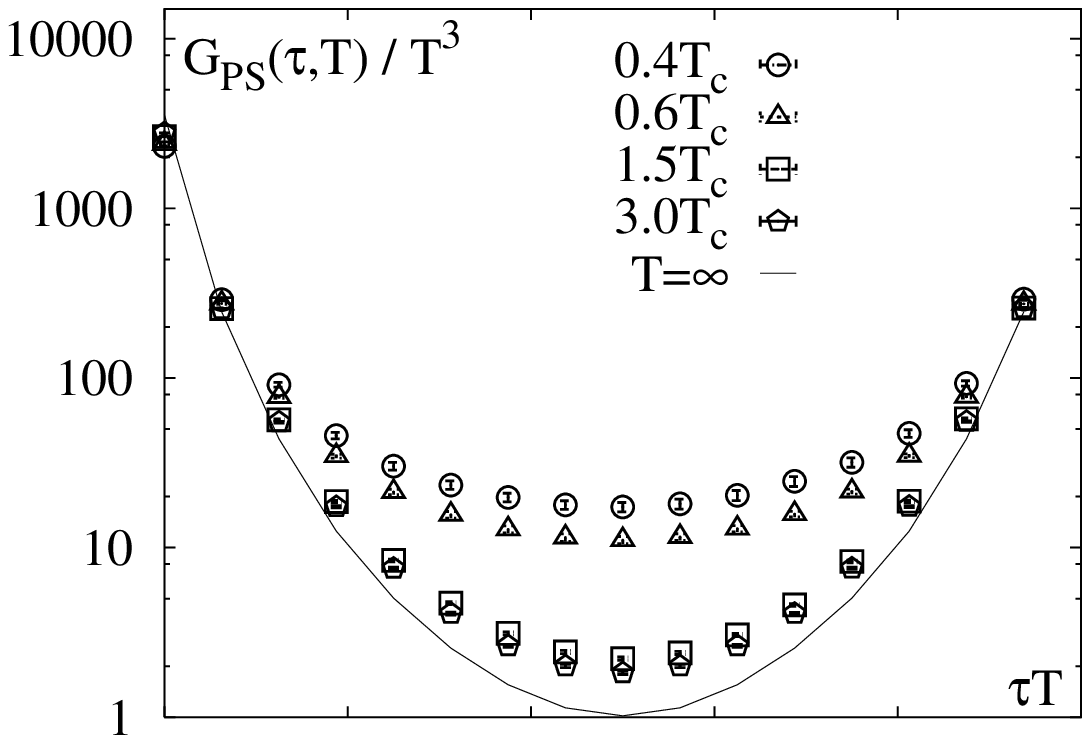}
\epsfxsize=5.6cm\epsfysize=3.6cm\epsfbox{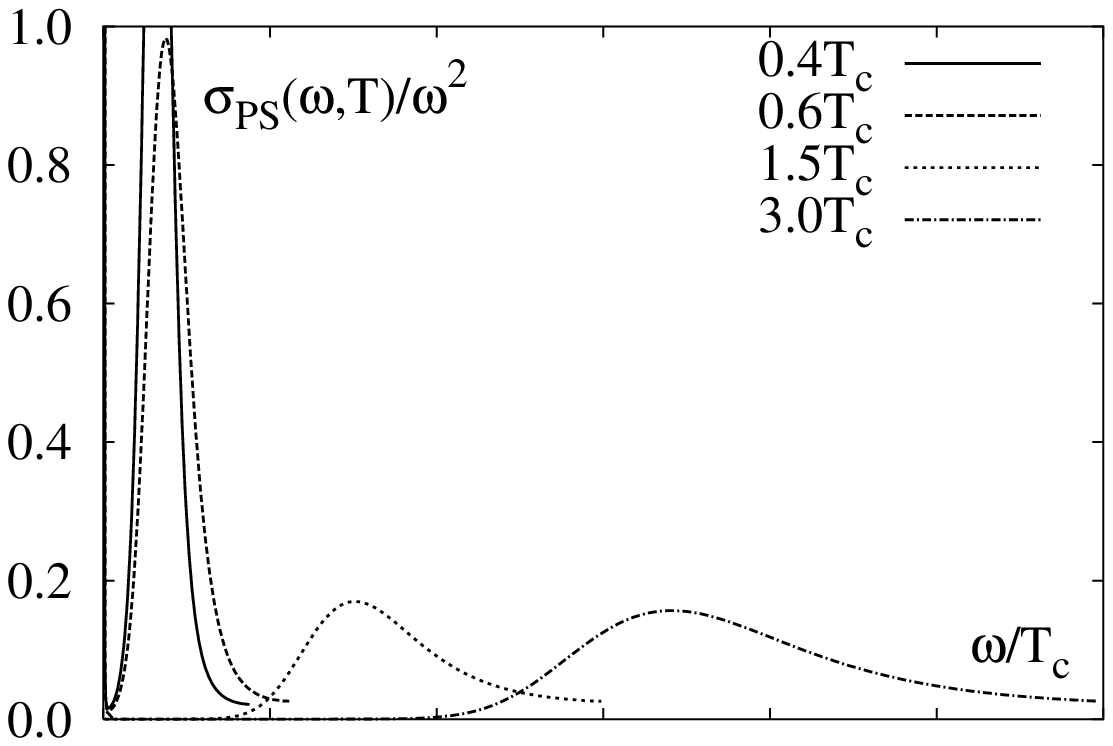}~\\
\epsfxsize=5.6cm\epsfysize=3.6cm\epsfbox{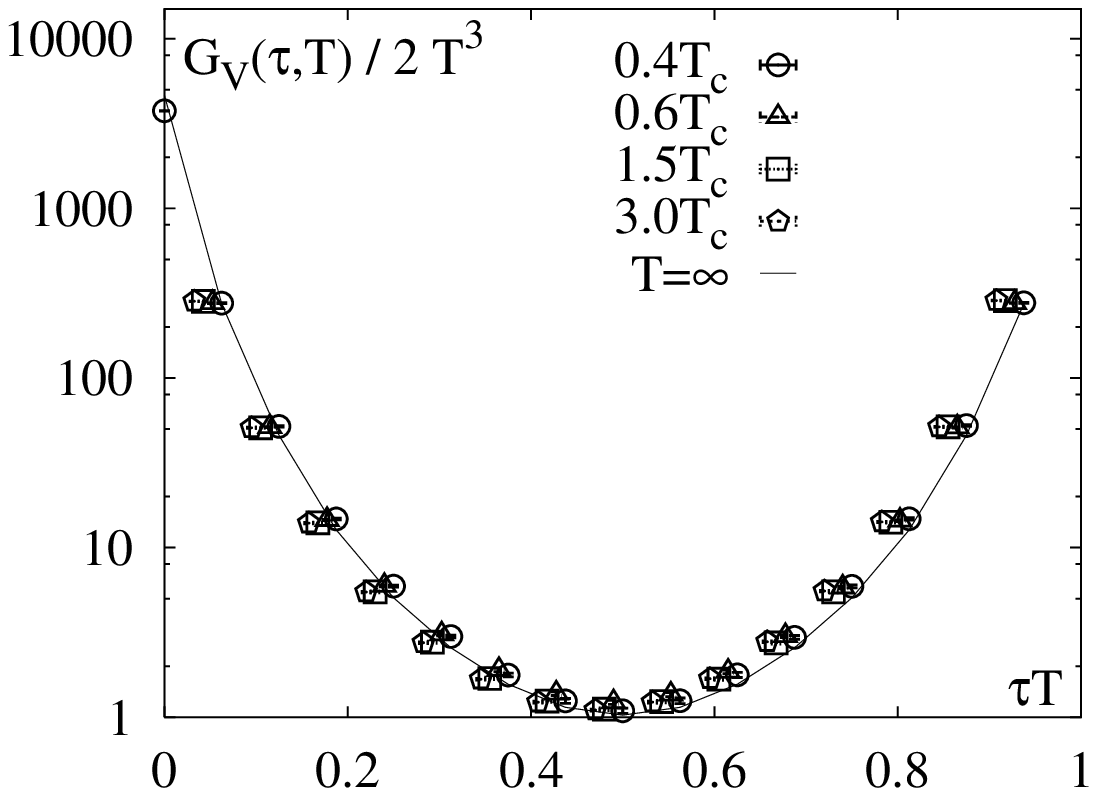}
\epsfxsize=5.6cm\epsfysize=3.6cm\epsfbox{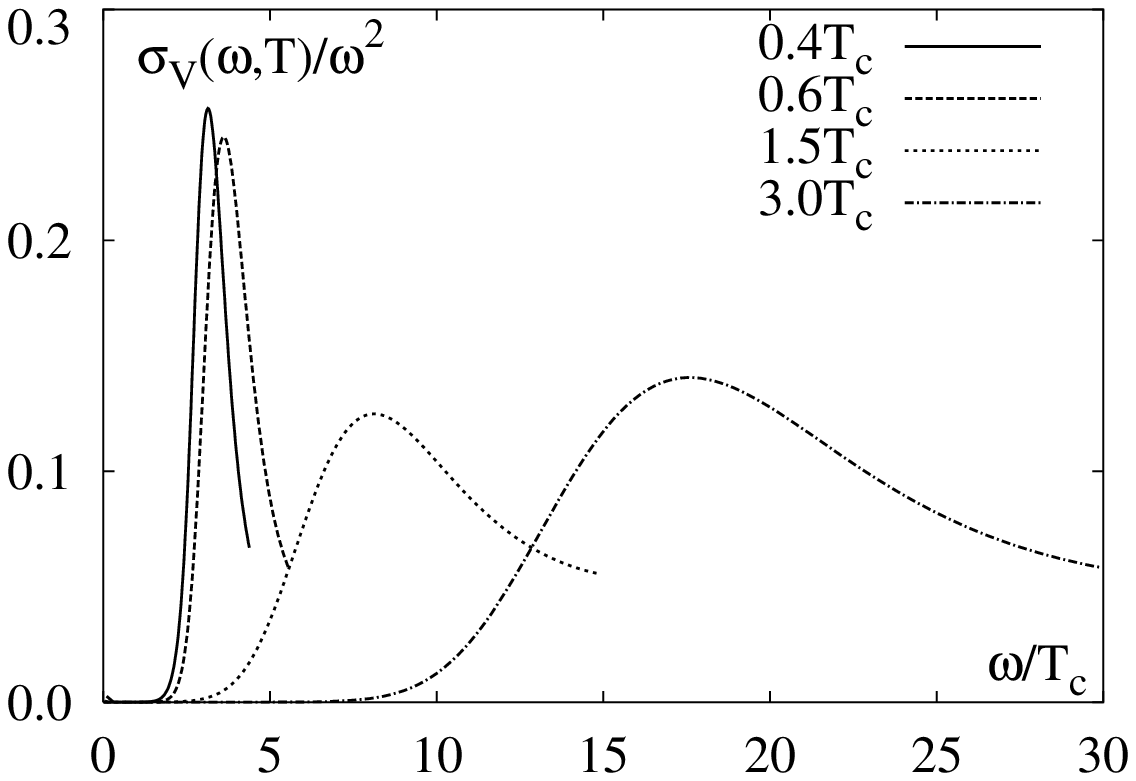}   
\caption{Light pseudo-scalar (top) and vector meson (bottom) correlation
(left) and spectral functions (right) in the range $0.4$ to $3.0 \; T_c$
taken from \protect\cite{qm02}. Points of the vector meson correlators 
are slightly displaced for better visibility. Only the low energy part
of the spectral functions is given, which is not affected by cut-off effects
\protect\cite{Stickan}.}
\label{fig_light}
\end{figure}

Figure \ref{fig_light} shows the modifications of the zero momentum correlation
and reconstructed spectral functions in the investigated temperature
range. While crossing the phase transition temperature the change of the
pseudo-scalar correlator is clearly obvious. Above the critical temperature the
correlation function slowly approaches the free meson correlator indicated as
solid line, while the vector meson correlator is near the free curve for all
temperatures. The corresponding spectral functions show a pronounced ground
state peak below $T_c$, which is transformed into a broad bump at higher 
energies above $T_c$. In fact, this remnant of the ground state peak
scales with the temperature and thus its position remains almost the same when
plotted in units of $\om/T$ instead of $\om/T_c$.

Moreover, a degeneracy of the pseudo-scalar and scalar iso-vector correlation
and spectral functions could be observed above the critical temperature
indicating effective chiral $U_A(1)$ symmetry restoration. 

\begin{figure}[ht]~\\
\epsfxsize=5.6cm\epsfysize=3.6cm\epsfbox{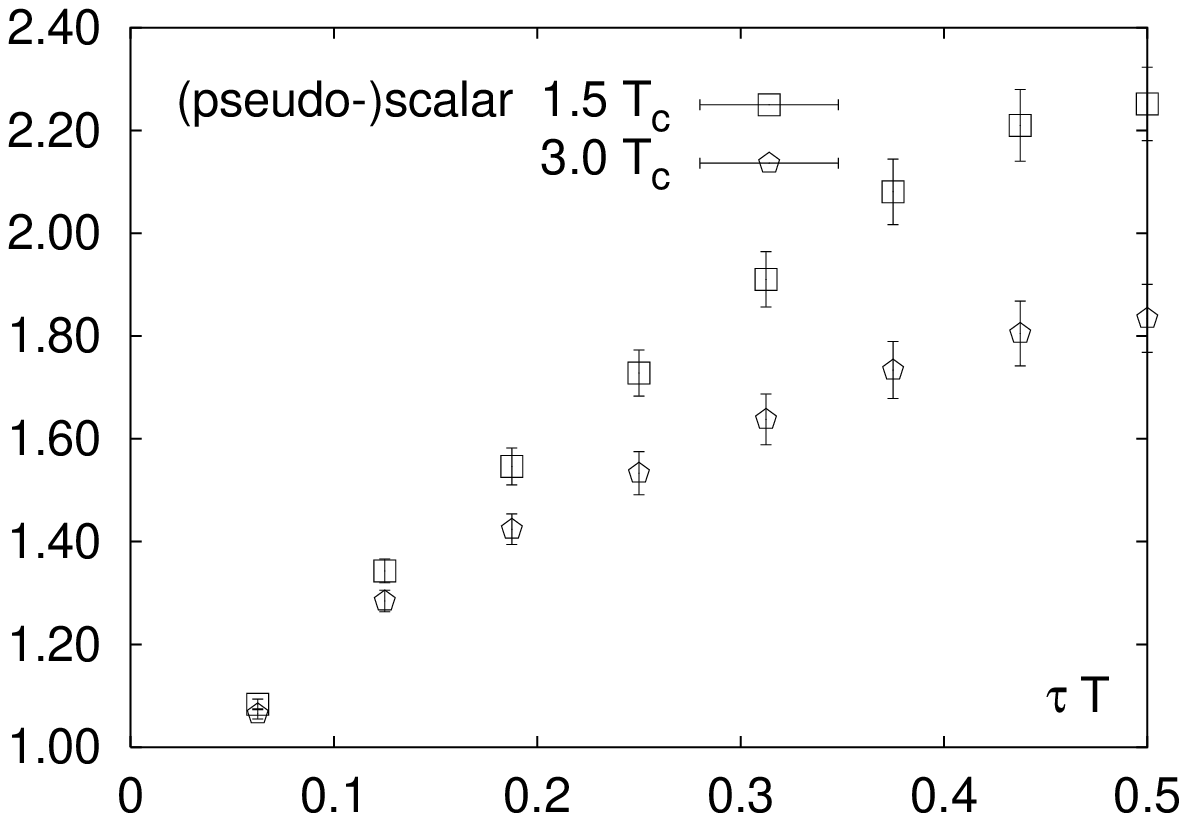}
\epsfxsize=5.6cm\epsfysize=3.6cm\epsfbox{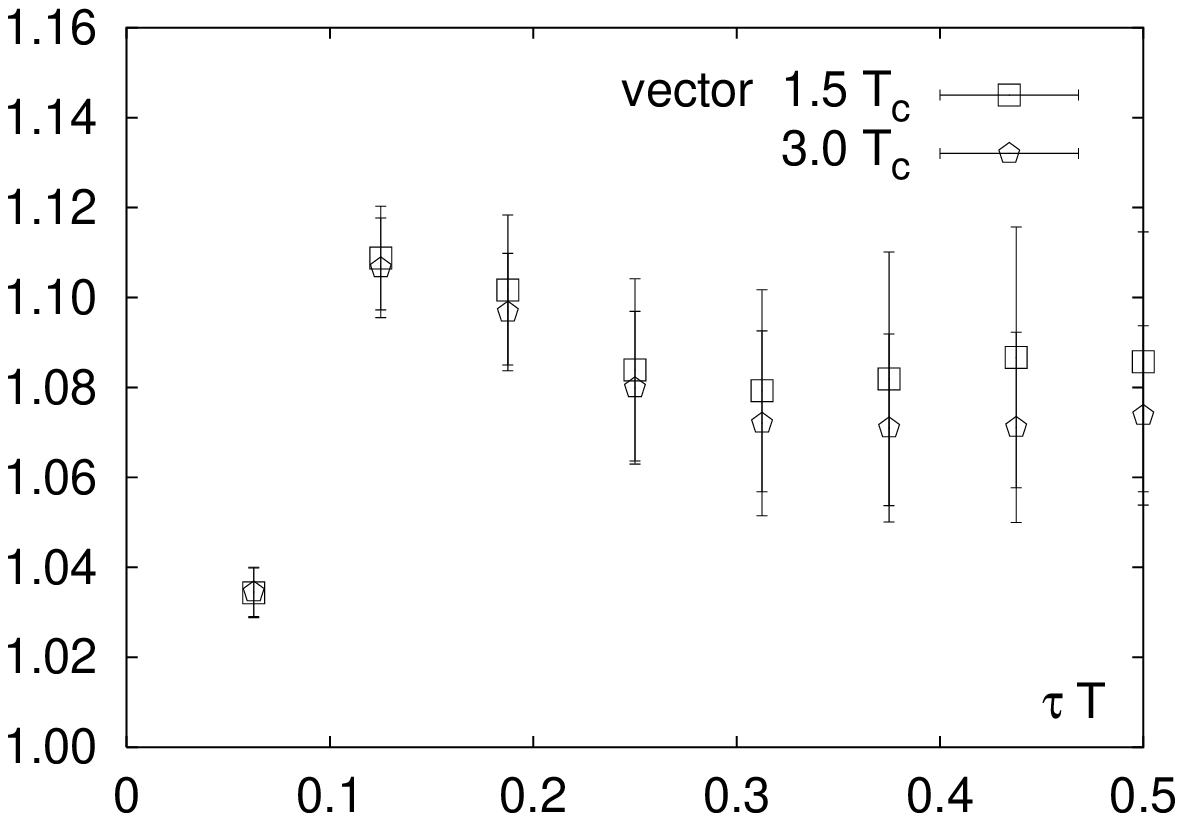}
\caption{Correlator ratios $G(\ta T)/G_{free}(\ta T)$
for the pseudo-scalar (left) and vector meson (right) taken from
\protect\cite{Dilept}.}
\label{fig_ratio}
\end{figure}

A more quantitative picture of the deviations from the free quark behavior
can be obtained by taking the ratio of the correlator with the
corresponding free meson correlator for the same lattice size, which is
illustrated in figure \ref{fig_ratio}. In the pseudo-scalar channel one
can still observe a considerable deviation from the propagation of a
free $q\bar{q}$ pair even at a temperature as high as 3 $T_c$ indicating 
a remnant of the strong interaction between the quarks above the chiral
phase transition. The situation is different in the vector channel. The
correlator ratio visualizes that the vector meson correlation function
shows only a 10 \% deviation from the free quark behavior. Furthermore, the
value of $G_V(1/2T,T)/T^3$ is finite, which implies that the spectral
function $\si_V$ has to vanish in the limit $\om \to 0$. The determination of
the slope of the spectral function at $\om=0$ would be of major interest, since
finite values for the transport coefficients of the QGP would require
$\si \sim \om$ in this limit \cite{Aarts}. Unfortunately, simulations on even
larger lattices would be needed to increase the sensitivity of the
reconstructed spectral functions in the very low energy region. 
 
\section{Thermal dilepton rates}
\begin{figure}[ht]~\\
\epsfxsize=5.6cm\epsfysize=3.6cm\epsfbox{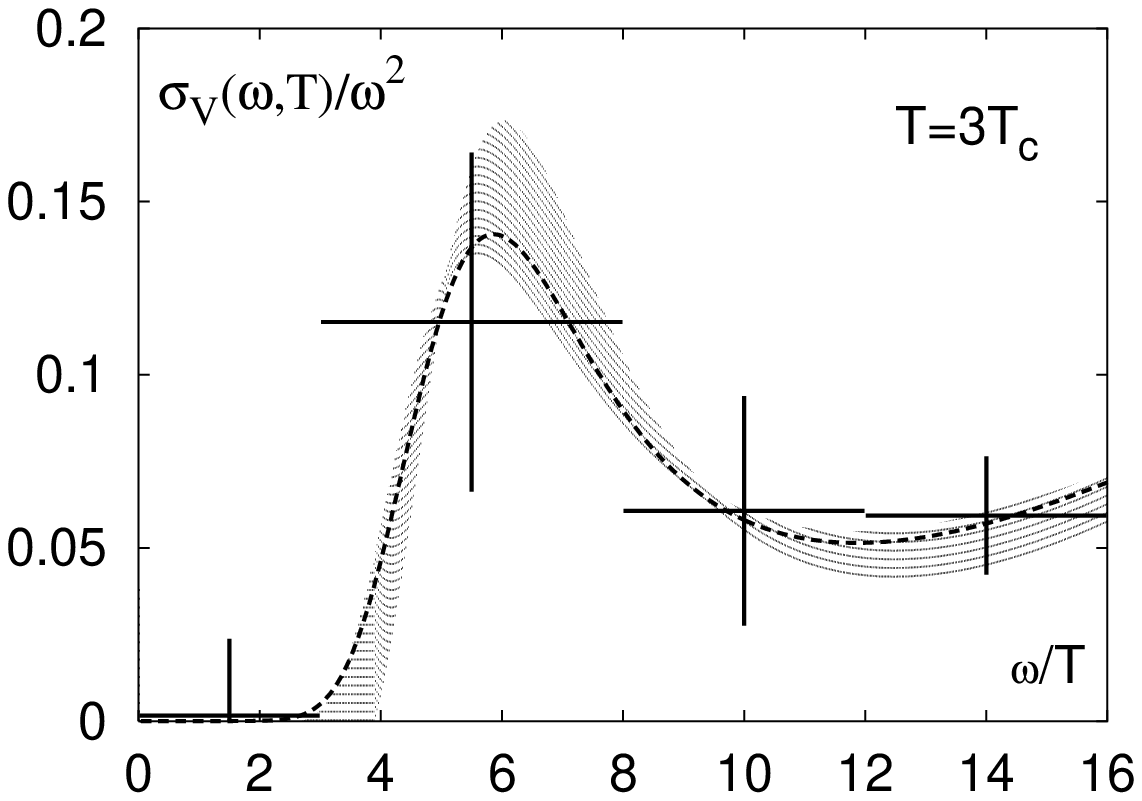}
\epsfxsize=5.6cm\epsfysize=3.6cm\epsfbox{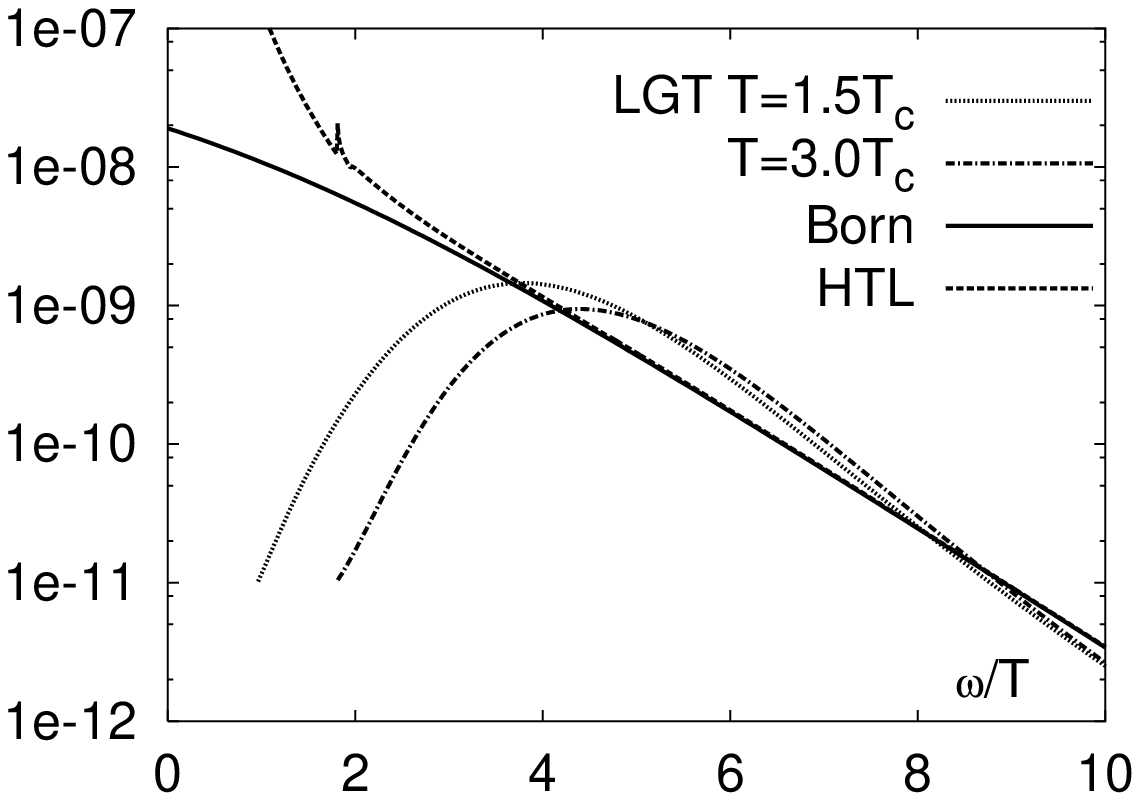}
\caption{Vector meson spectral function (left) and thermal
dilepton rate at $p=0$ (right) taken from
\protect\cite{Dilept,qm02}.}
\label{fig_dilept}
\end{figure}

The thermal dilepton spectrum is accessible in heavy ion collisions and
provides an important observable to study thermal properties of the medium
at high temperature and density \cite{Kapusta,Toimela_Alam}. The
non-perturbative in-medium modifications of the $q\bar{q}$ interactions are
expected to influence thermally induced changes of the dilepton spectrum
at low energies \cite{Abreu}. The differential dilepton rate in two-flavor
QCD is directly related to the spectral function in the vector channel
\bqa
\frac{dN_{l\bar l}}{d^4xd^4p} \;\;\equiv\;\;
\frac {dW} {d\omega d^3p} &=& 
\frac {5\alpha^2} {27\pi^2} \;
\frac {1} {\omega^2 \left({ e^{\om/T} -1 }\right)} \;
\sigma_V(\omega,\vec{p},T) \;.
\label{rate}
\eqa
The left hand side of figure \ref{fig_dilept} gives a detailed error estimate
on the vector meson spectral function which is used to calculate the dilepton
rate. The dashed band illustrates solely the statistical error, while the
vertical lines illustrate the uncertainty of the MEM reconstruction in the
given $\om/T$ ranges. Moreover, calculations with different lattice spacings
and volumes have been performed to ensure that the spectral shape of $\si_V$ is
not influenced by cut-off or finite volume effects in the given energy regime
\cite{Stickan}.

The thermal cross section for the production of dilepton pairs at vanishing
momentum (right side of fig.\ref{fig_dilept}) is obtained by applying
eq.(\ref{rate}) for the vector meson spectral function. The leading
order perturbative result (Born rate) and a calculation in HTL-resummed
perturbation theory \cite{htl,Mustafa} are given for comparison. A small
enhancement over the Born rate in the range 4-8 $\om/T$ is visible for the
lattice gauge theory (LGT) result at both temperatures, whereas the rate
decreases rapidly at smaller energies. This behavior is in clear contrast to
the hard thermal loop calculations, which predict a divergent result in the
infrared limit.

It is evident that perturbative as well as lattice calculations
should investigate the low energy regime in more detail. An obvious step in
this direction is to try to enhance the sensitivity of the statistical analysis
of thermal correlation functions in the low energy region. A first attempt
along this line has recently been tested in a calculation of the electrical
conductivity and soft photon production rate of the QCD plasma \cite{Gupta}.

\section{Charmonia}
For the heavy $c\bar{c}$ bound states the situation is different from
the light quark sector. Potential model calculations predict a sequential
dissolution, where higher excitations like $\chi_{c,0}\;[^3P_0]$ and
$\chi_{c,1}\;[^3P_1]$ dissolve earlier, while the s-wave states $J/\psi\;
[^3S_1]$ and $\eta_c\;[^1S_0]$ may survive after the deconfinement phase
transition \cite{Satz}.

\begin{figure}[ht]~\\
\epsfxsize=5.6cm\epsfbox{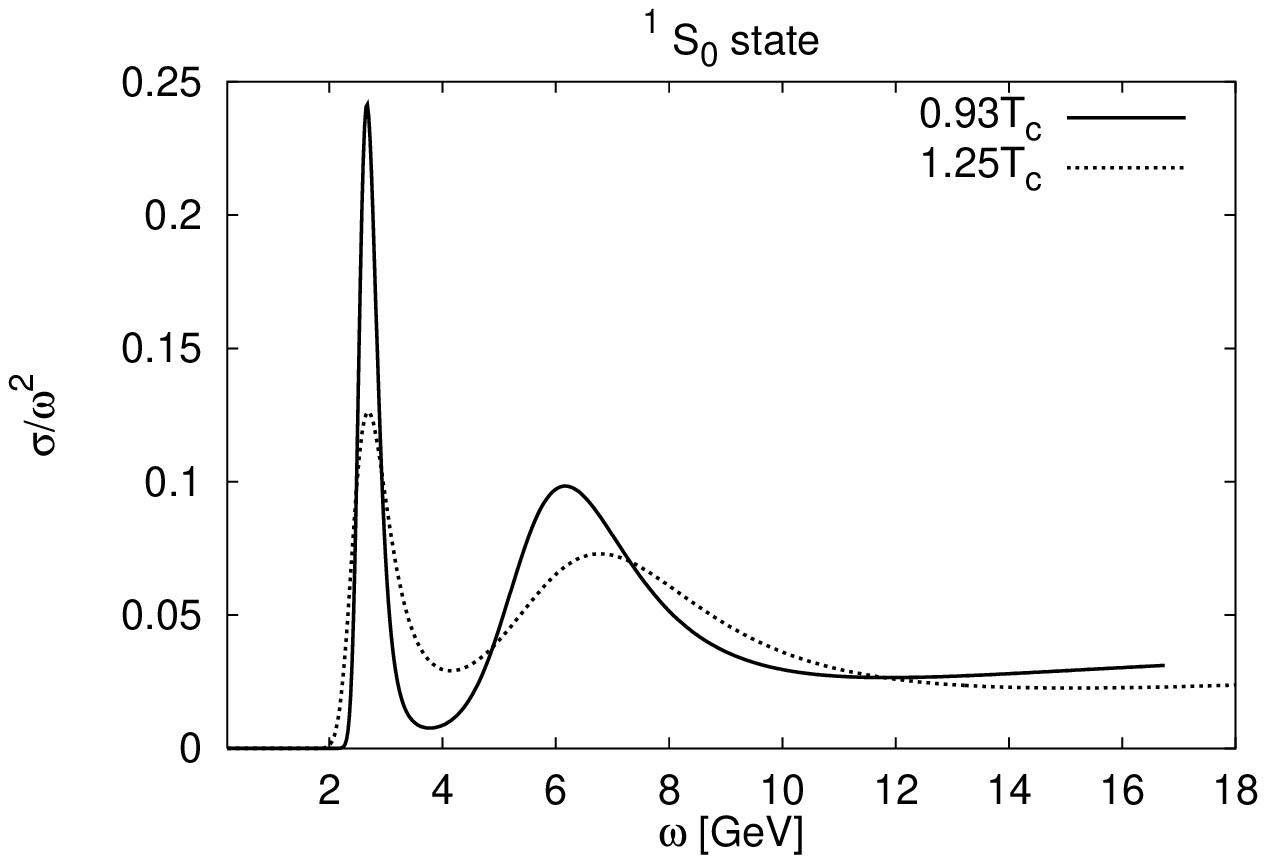}
\hspace*{2mm}
\epsfxsize=5.6cm\epsfbox{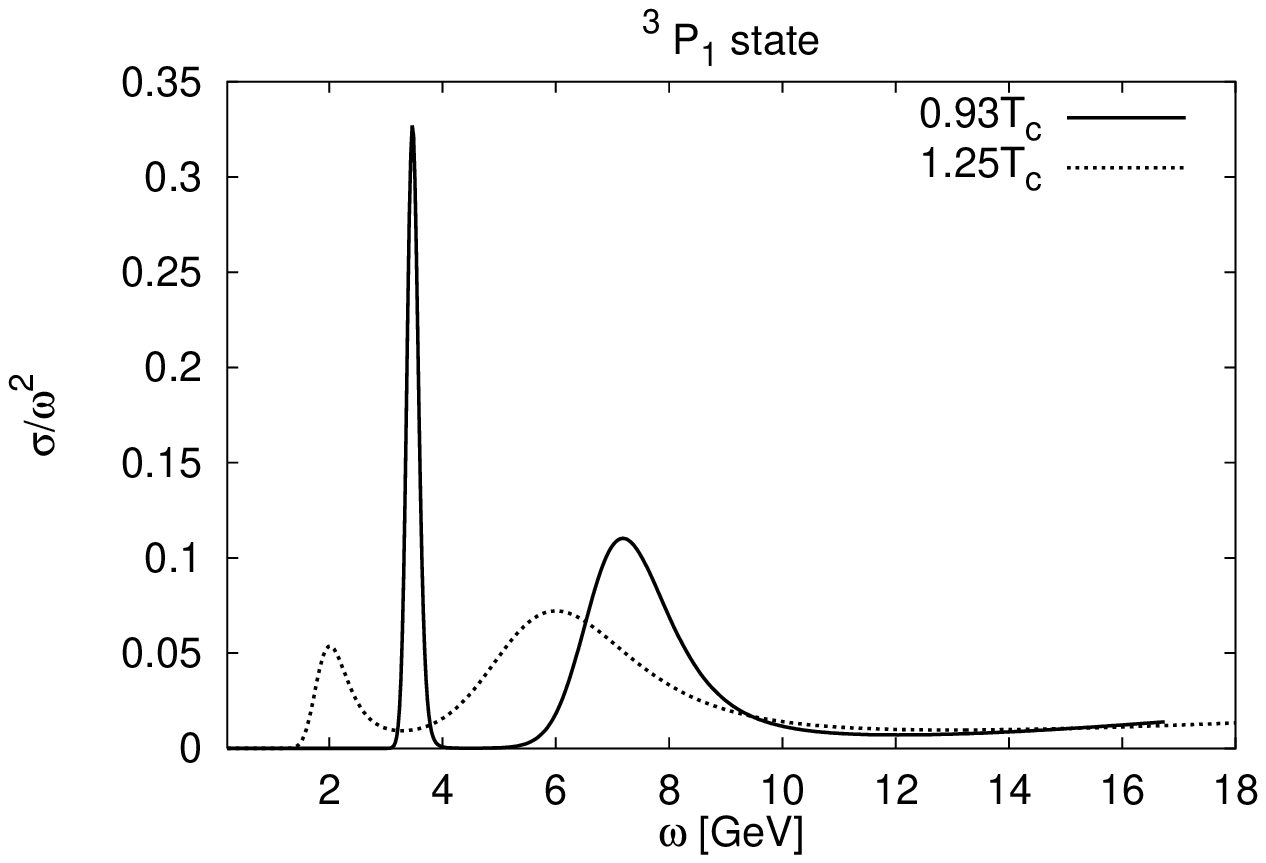}   
\caption{Thermal changes of charmonium spectral functions in the
pseudo-scalar (left) and axial-vector channel (right) taken from
\protect\cite{Datta}.}
\label{fig_charm}
\end{figure}

Three different lattice QCD groups have investigated charmonium
spectral functions so far. The simulations have been performed on
large isotropic \cite{Datta} and anisotropic lattices \cite{Umeda,Asakawa}
covering the temperature range 0.9 to 1.9 $T_c$. The thermal modifications
of the charmonium spectral functions found by \cite{Datta} are displayed in
figure \ref{fig_charm} for the pseudo-scalar ($\eta_c$ and similar the vector
meson $J/\psi$) and for the axial-vector channel ($\chi_{c,1}$ and similar for
the scalar $\chi_{c,0}$) at 0.93 and 1.25 $T_c$. The left figure shows that the
pronounced ground state peaks for the $J/\psi$ and $\eta_c$ survive even at
1.25 $T_c$, while a drastic change can be observed in the spectral functions of
the orbitally excited states $\chi_{c,0}$ and $\chi_{c,1}$. The second broad
bump at higher energies in both pictures is most likely a cut-off effect as
discussed above. 

The results on anisotropic lattices support this finding of sequential
dissolution. The survival of sharp ground state peaks at 1.1 $T_c$ has been
reported for the pseudo-scalar ($\eta_c$) and vector channel ($J/\psi$)
\cite{Umeda}, while a near degeneracy of the spectral functions in all channels
has been observed at 1.9 $T_c$ indicating chiral symmetry restoration
\cite{Asakawa}. Combining the results of all groups one may conclude that the
orbitally excited states dissolve in the vicinity of the deconfinement phase
transition temperature, while the $J/\psi$ and $\eta_c$ persist as bound states
up to 1.25 $T_c$ and probably dissolve in the range 1.5 - 1.9 $T_c$.

\section{Conclusions \& Outlook}
The application of the Maximum Entropy Method for the reconstruction
of spectral functions from correlation functions in Euclidean time
without a priori assumptions on the spectral shape initiated the recent
progress of lattice investigations in thermal physics. Already at zero
temperature this approach leads to more precise predictions of the masses of
excited states, but the most important point is that the study of hadronic
spectral functions at finite temperature became feasible. New insight
could be obtained in the in-medium modifications of hadronic
properties in the vicinity and above the phase transition temperature.

Deviations from the free quark-antiquark propagation could be observed in the
pseudo-scalar channel up to 3 $T_c$, while the vector meson correlation and
spectral function are much closer to the free quark behavior. Above the
critical temperature an effective $U_A(1)$ symmetry restoration between
the pseudo-scalar and scalar iso-vector states has been found. The thermal
dilepton production rate could be calculated for the first time from lattice
QCD data using the spectral function in the vector meson channel.
Finally, the pattern of sequential dissolution of the heavy charmonium states
is supported by the spectral analysis.

Further investigations will be certainly needed in the vicinity of the critical
temperature and in the low energy region. All the lattice QCD calculations
reported in this summary were performed in the quenched approximation, which
neglects the influence of virtual quark loops. The extension to simulations
with dynamical quarks will be the major challenge in the future.

\section*{Acknowledgments}
Parts of the work summarized in this talk emanate from a very lively
collaboration with F.~Karsch, E.~Laermann, S.~Datta, P.~Petreczky and
S.~Stickan.
Furthermore, I want to thank the organizers of the Villefranche Workshop on
Quantum Chromodynamics for the invitation.

\end{document}